\documentclass[amsmath,amssymb,aps,prc,twocolumn,superscriptaddress,showpacs,preprintnumbers]{revtex4}

\usepackage{graphicx,color}
\usepackage{multirow}
\usepackage{times}
\usepackage{bm}
\usepackage{epstopdf}

\usepackage[normalem]{ulem}  

\newcommand{\comment}[1]{}
\renewcommand\sout{\bgroup \color{red} \ULdepth=-.5ex \ULset}
\newcommand{\srtNN}{\sqrt{s_{{\scriptscriptstyle NN}}}}

\begin{document}

\title{
Equation of state dependence of directed flow
in a microscopic transport model
}
\author{Yasushi Nara}
\affiliation{
Akita International University, Yuwa, Akita-city 010-1292, Japan}
\affiliation{Frankfurt Institute for Advanced Studies, 
D-60438 Frankfurt am Main, Germany}
\author{Harri Niemi}
\affiliation{Institut f\"ur Theoretishe Physik,
 Johann Wolfgang Goethe Universit\"at, D-60438 Frankfurt am Main, Germany}
%
\author{Jan Steinheimer}
\affiliation{Frankfurt Institute for Advanced Studies, 
D-60438 Frankfurt am Main, Germany}
\author{Horst St\"ocker}
\affiliation{Frankfurt Institute for Advanced Studies, 
D-60438 Frankfurt am Main, Germany}
\affiliation{Institut f\"ur Theoretishe Physik,
 Johann Wolfgang Goethe Universit\"at, D-60438 Frankfurt am Main, Germany}
\affiliation{GSI Helmholtzzentrum f\"ur Schwerionenforschung GmbH, D-64291
Darmstadt, Germany}

\date{\today}

\begin{abstract}
We study the sensitivities of the directed flow in Au+Au collisions 
on the equation of state (EoS), employing the transport theoretical model JAM.
The EoS is modified by introducing a new collision term in order to control the pressure of a system
by appropriately selecting an azimuthal angle in two-body collisions
according to a given EoS.
It is shown that this approach is an efficient method to modify the EoS in a transport model. 
The beam energy dependence of the
directed flow of protons is examined with two different EoS,
a first-order phase transition and crossover.
It is found that our approach yields quite similar results as hydrodynamical
predictions on the beam energy dependence of the directed flow;
Transport theory predicts a minimum in the excitation function of
the slope of proton directed flow and does indeed yield negative directed flow,
if the EoS with a first-order phase transition is employed.
Our result strongly suggests that the highest sensitivity for
the critical point can be seen in the beam energy range of
$4.7\leq\srtNN\leq11.5$ GeV.

\end{abstract}

\maketitle

One of the most challenging problems in high energy heavy ion collisions
is to map out the QCD phase diagram from low to high baryon densities.
In particular, the main interest is to discover
a first and/or second order
phase transition together with the critical point of QCD matter
at finite baryon chemical potentials,
which was predicted by several effective models
based on QCD~\cite{PhaseDiagram}.
In order to explore QCD matter in the full range of temperatures
and baryon densities,
it is necessary to measure various observables
for a wide range of beam energies.
A non-trivial beam energy dependence of various observables such as the
$K^+/\pi^+$ ratio~\cite{Alt:2007aa},
higher order of the net-proton number cumulants
~\cite{Adamczyk:2013dal,Luo:2015doi} and
the slope of the directed flow have been reported from the NA49 Collaboration
and the STAR beam energy scan (BES) program
~\cite{Aggarwal:2010cw,Kumar:2013cqa}.
We note that such a beam energy dependence usually is not explained by the
standard hadronic transport models. There have been attempts
to take into account a change of degrees of freedom (into quarks and gluons)
~\cite{AMPT,PHSD} in transport models,
or transport + hydrodynamics hybrid models~\cite{Petersen:2008dd}.
For instance, hybrid models usually show an improved description of strange particle
ratios and collective flow~\cite{Steinheimer:2011mp}

The collective transverse flow developed in the heavy ion collisions
is considered to be sensitive to the equation of state (EoS) of QCD matter,
since it reflects the pressure gradients in the early stages of the reaction.
Especially, the softest region in the EoS with a (first-order) phase transition
is expected to have significant influence on the directed flow
$v_1=\langle \cos\phi\rangle$
of nucleons~\cite{Rischke:1995pe,Stoecker:2004qu}.
Hydrodynamic predictions have shown that the
excitation function of directed flow exhibits a local minimum at
a specific beam energy characteristic for the first-order phase transition
~\cite{Rischke:1995pe,Brachmann:1999xt,Ivanov:2000dr}.
Furthermore, the slope of the directed flow of nucleons becomes
negative at the softest point of the EoS.~\cite{Csernai:1999nf,Csernai:2004gk,
Brachmann:1999xt,Brachmann:1999mp}.
Thus the collapse of directed flow may be a signature of
the QCD first-order phase transition at high baryon densities.
Indeed, such a behavior has been observed in the BES program
by the STAR Collaboration~\cite{STARv1,Shanmuganathan:2015qxb,Singha:2016mna}
but no microscopic hadronic transport model is yet able to quantitatively explain the negative
slope of protons at $\srtNN=11.5$ and 19.6 GeV
~\cite{Konchakovski:2014gda,Petersen:2006vm,Steinheimer:2014pfa}.
Only at higher beam energies,
the microscopic transport models RQMD~\cite{Snellings:1999bt},
UrQMD~\cite{Bleicher:2000sx},
and PHSD/HSD~\cite{Konchakovski:2014gda}
show a negative slope of proton directed flow
without a phase transition.
We would like to note that studies including a hadronic mean field
interaction also have not been able to describe the negative slope
of the proton directed flow~\cite{Isse,qm15no}.
The excitation function of the directed flow
has also been examined by other approaches such as
the UrQMD hybrid model~\cite{Steinheimer:2014pfa},
three fluid models~\cite{Ivanov:2014ioa}, 
and JAM with an attractive scattering style~\cite{Nara:2016phs},
but none of the models is able to give a quantitative satisfactory description
of the STAR data. 
Therefore it is still premature to make unambiguous interpretations
of the STAR data, and more reliable models have to be developed in order to
simulate the spacetime evolution of the system created in heavy ion
collisions at intermediate beam energies and to understand the underlying collision dynamics.

There is a long history of implementing an EoS into transport theoretical
approaches. The Boltzmann-Uehling-Uhlenbeck (BUU) model~\cite{Bertsch:1984gb}
was developed first followed by quantum molecular dynamics
(QMD)~\cite{Aichelin:1986wa} to test various EoS in a transport model
that includes nuclear mean field potentials.
Later a relativistic extension of both BUU and QMD 
was developed which are called RBUU~\cite{RBUU} and RQMD~\cite{RQMD}.
It is also known that the EoS can be controlled by changing
the scattering style in the two-body collision term
~\cite{Halbert:1981zz,Gyulassy:1981nq,Kahana:1994be}.
For example, choosing a repulsive orbit in two-body collisions
can simulate the effect of a repulsive potential.
Later, effects of a phase transition have been investigated
by including a mean field interaction~\cite{Li:1998ze,Danielewicz:1998pb}
or using attractive orbits
in two-body collisions~\cite{Sorge:1998mk,Nara:2016phs}.

The theoretically favored method is to include mean fields to vary the EoS,
and in principle any mean field can be incorporated
provided that the interaction Lagrangian is known.
It is, however, practically very challenging to implement a mean field in a dynamical simulation,
and one often relies on the simplified phenomenological form of potentials.
On the other hand, an advantage of the second approach is that 
we can include any kind of EoS computed by other methods
such as effective models or lattice QCD.
In this work, we shall follow the second approach to examine
effects of the EoS on the directed flow
within the microscopic transport approach JAM by modifying the scattering style
in the two-body collision term. 
We do not include explicitly partonic degree of freedom,
as for example in AMPT~\cite{AMPT} or PHSD~\cite{PHSD},
since the transition from one set of degrees of freedom is difficult
and usually introduced new uncertainties in the model.
See also Ref.~\cite{Feng:2016ddr}
for a recent attempt to implement a first order
phase transition into a parton cascade model.

In JAM~\cite{JAM}, particle production is modeled by the excitation and
decay of resonances and strings
similar to the other models~\cite{RQMD1995,UrQMD1,UrQMD2}.
Secondary products from decays can interact with each other
via binary collisions.
A detailed description of the JAM model can be found
in Ref.~\cite{JAM,Hirano:2012yy}.

In the standard cascade approach the two body collision term is implemented
so that it does not generate additional pressure.
Namely, the azimuthal angle in the two-body collision is randomly chosen.
It is well known that the pressure can be controlled by changing
the scattering style.
Here we employ a method similar to that proposed in Ref.~\cite{Sorge:1998mk}
in which the momentum change in each two-body collision between
particle $i$ and $j$ at the space-time coordinates of $q_i$ and $q_j$
is related to the pressure $\Delta P=P-P_f$, where $P_f$ is the free
streaming part of the local pressure.
Motivated by the virial theorem~\cite{Danielewicz:1995ay},
The formula is given by 
\begin{equation}
  \Delta P = -\frac{\rho}{3(\delta\tau_i+\delta\tau_j)}
                 (p_i'-p_i)^\mu (q_i-q_j)_\mu,
\label{eq:pre}
\end{equation}
where $\rho$ is the Lorentz invariant local particle density
obtained by $\rho=N^\nu u_\nu$,
$N^\nu=\sum_h\int\frac{d^3p}{p^0}p^\nu f_h(x,p)$,
$\delta\tau_{i}$ is the proper time interval between
successive collisions, and $p'_i-p_i$ is
the energy-momentum change of the particle $i$.
$P_f$ can be computed from the energy-momentum tensor $T^{\mu\nu}$:
$P_f=-\frac{1}{3}\Delta_{\mu\nu}T^{\mu\nu}$,
where
$T^{\mu\nu}=\sum_h\int \frac{d^3p}{p^0}p^{\mu}p^{\nu}f_h(x,p)$
with the projector of $\Delta^{\mu\nu}=g^{\mu\nu}-u^\mu u^\nu$,
where $u^\nu$ is a hydrodynamics velocity defined
by the Landau and Lifshitz's definition.
In this way, we control the pressure such that
$P$ coincides with a given EoS at an energy density $e=u_\mu T^{\mu\nu}u_\nu$.

There are three ways to satisfy the constraint Eq.(\ref{eq:pre})
to control the pressure in two-body collisions;
\begin{enumerate}
\item change the scattering angle,
\item change the azimuthal angle,
\item change the magnitude of the relative momentum by modifying the masses
of the outgoing particles.
\end{enumerate}
In Ref.~\cite{Sorge:1998mk}, an additional elastic scattering is
generated after any of the standard two-body scatterings, and it was found that
more elliptic flow is generated.
It is known that the shear viscosity is sensitive to
the scattering angle~\cite{Wesp:2011yy}.
Thus, it is unclear whether the EoS alone is responsible for the
modification of elliptic flow, since transport coefficients are also
modified at the same time by additional extra elastic scatterings.
We therefore avoid changing the scattering angle in this work in order to
model the EoS and effectively keeping the transport coefficients
as unchanged as possible.
The idea of changing the outgoing mass is attractive in the sense that
it may effectively simulate a dropping mass due to
chiral symmetry restoration (CSR).
But it is not straightforward how to relate it to CSR within
our approach and hadrons would eventually have to retain their vacuum masses before escaping the dense system.
Thus, in this work, we adapt the method of only changing
the azimuthal angle in the two-body collision so that the constraint
Eq.(\ref{eq:pre}) is fulfilled.

We test two types of EoS. For the first-order phase transition,
the hadron resonance gas phase contains all hadronic resonances up to 2 GeV,
and a baryon density $\rho_B$ dependent vector type repulsive mean field
$V(\rho_B)=\frac{1}{2}K\rho_B^2$ with $K=0.45$
GeV fm$^3$ is included as described in ~\cite{EOSQ}.
The QGP phase is modeled by the ideal gas of massless quarks and gluons with a 
bag constant of $B^{1/4}=220$ MeV which leads to a phase transition
temperature of $T_c=156$ MeV at vanishing net baryon density.
For the crossover EoS, we use the chiral model EoS from Ref.~\cite{CMEOS}
in which EoS at vanishing and finite baryon density is consistent with a smooth crossover transition,
as found in lattice QCD results.

\begin{figure}[tbh]
\includegraphics[width=8.5cm]{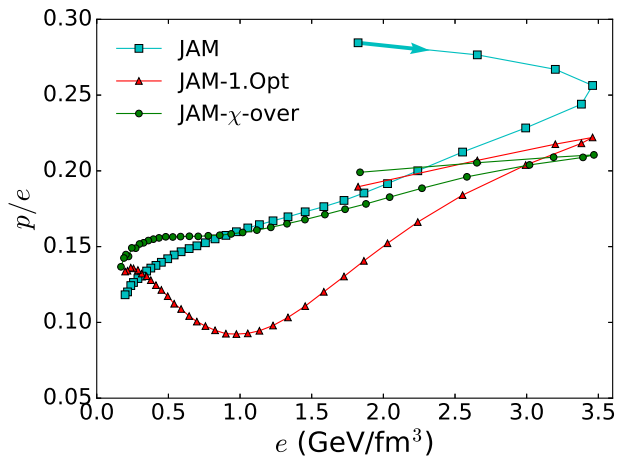}
\caption{Time evolution of isotropic pressure $p$ and energy density $e$
in mid-central Au+Au collisions (10-40\%)
at $\srtNN=11.5$ GeV
from JAM cascade mode (squares),
JAM with first-order EoS (triangles)
and with crossover EoS (circles).
Initial points corresponds to the time 0.65 fm/c after touching
of two nuclei, and time interval is 0.25 fm/c.
Arrow indicates the direction of the reaction.
Pressure and energy density
are averaged over collision points in a cylindrical volume of transverse
radius 3 fm and longitudinal distance of 3 fm centered at the origin.
}
\label{fig:eos11}
\end{figure}

In Fig.~\ref{fig:eos11}, the time evolution of the local
isotropic pressure and energy density, averaged over events,
in mid-central Au+Au collision at $\srtNN=11.5$ GeV
are compared for the different EoS used.
The JAM standard EoS is the stiffest, which is due to 
the chemical non-equilibrium in the initial pre-equilibrium evolution 
as pointed out in Ref.~\cite{Nara:2016phs}.
A pronounced softening can be seen in the case of the first order phase
transition at around the energy density of $e\approx 1$ GeV/fm$^3$
as we expected.
On the other hand, the effect of softening is weaker for the crossover EoS.
In the construction of the crossover EoS from a chiral model,
correct nuclear saturation properties at zero temperature where required,
resulting in the stiff hadronic EoS, compared to the standard
JAM EoS which is expected to be close to an ideal hadron resonance gas EoS.
This property leads effectively to the inclusion of repulsive potential
in the hadronic phase in JAM simulations.

\begin{figure}[t]
\includegraphics[width=8.0cm]{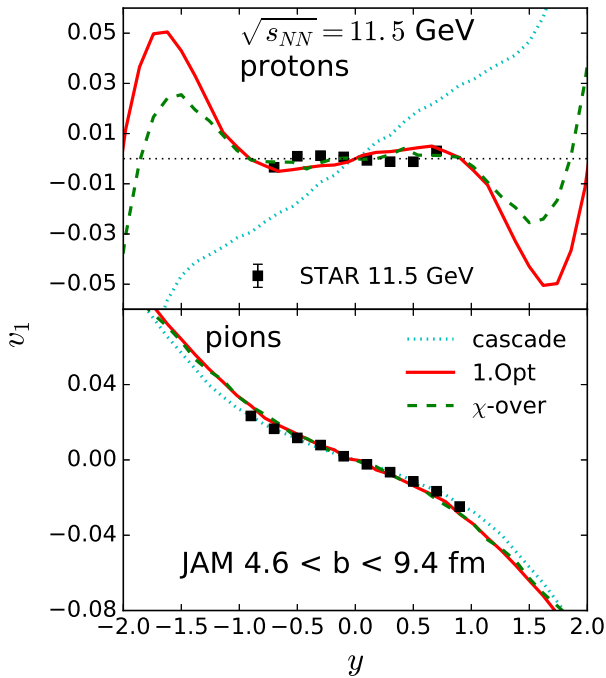}
\caption{Rapidity dependence of directed flow $v_1$ for protons and pions
in semi-central Au+Au collisions
for $\srtNN=11.5$ GeV from JAM simulations
with different EoSs.  The solid line presents the JAM with the EoS with 1st
order phase transition, and the dashed line presents the JAM with the
crossover  EoS.  The dotted line is for the standard JAM result.
The STAR data are taken from Ref.~\cite{STARv1}.
}
\label{fig:v1_11}
\end{figure}

We now compute the directed flow in mid-central Au+Au collisions, and
compare the two EoS described above.
In the simulation, we choose the impact parameter range
$4.6<b<9.4$ fm for mid-central collisions.
In Fig.~\ref{fig:v1_11}, we show the rapidity dependence 
of the proton (upper panel) and pion (lower panel)
directed flows in mid-central Au+Au collision
at $\srtNN=11.5$ GeV.
The pion directed flow is not sensitive to the EoS, all models are able reproduce the pion data,
which means the pion directed flow is not useful to probe the early time dynamics of the collision.
As we showed in Ref.~\cite{Nara:2016phs}, that the JAM cascade calculation
for protons shows a significantly larger flow than the STAR data.
It is seen that the proton directed flow from JAM with both the first order
phase transition and crossover EoS
are significantly reduced at mid-rapidity,
and close to the data, indicating the
importance of the EoS effects in the early phase.
However, both EoS still predict a positive $v_1$-slope contrary to
the STAR data.

\begin{figure}[t]
\includegraphics[width=8.0cm]{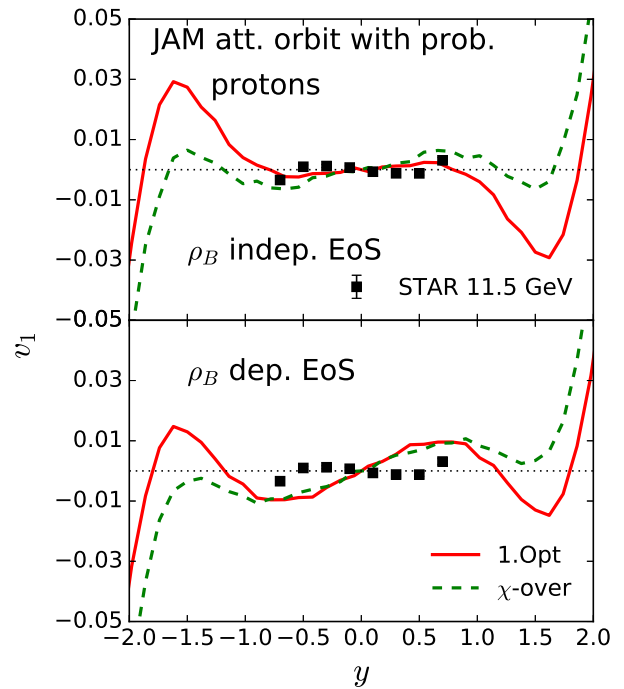}
\caption{Rapidity dependence of directed flow $v_1$ for protons
in semi-central Au+Au collisions
for $\srtNN=11.5$ GeV from JAM simulations by choosing attractive
orbit with a probability given by Eq.~(\ref{eq:attprob}).
The solid line presents the JAM with the EoS with 1st
order phase transition, and the dashed line presents the JAM with the
crossover  EoS. 
Upper panel: the results from baryon density independent EoS used
in Ref.~\cite{Nara:2016phs}.
lower panel: the results from baryon density dependent EoS.
The STAR data are taken from Ref.~\cite{STARv1}.
}
\label{fig:v1att11}
\end{figure}

We now check if the microscopic details are important for the generation
of directed flow.
In order to see systematics on the modeling in controlling a EoS,
we show in Fig.~\ref{fig:v1att11}, the rapidity dependence of
proton directed flow at $\srtNN=11.5$ GeV
from the method of choosing attractive
orbit in two-body scattering with the probability:
\begin{equation}
 p_\text{attractive} = \max\left(0,\frac{P_f - P(e)}{P(e)}\right),
 \label{eq:attprob}
\end{equation}
as proposed in Ref.~\cite{Nara:2016phs},
where $P(e)$ is a pressure as a function of energy density $e$ from a given
EoS.
As we simply choose attractive orbit according to the probability
given by Eq.~(\ref{eq:attprob}) in this method, pressure at
each two-body collision is not equal to the pressure of a given EoS,
and it fluctuates a lot. Only average pressure of the system
coincides with the given EoS. 
With this approach, one see that proton directed flow is less suppressed 
compared with our new method.
We also show the directed flow obtained by
the baryon density independent EoS as used in Ref.~\cite{Nara:2016phs}
in the upper panel of Fig.~\ref{fig:v1att11}.
Baryon density dependent EoS used in this work yields 
less suppression of the proton directed flow as we expected, 
since our EoS predicts less softening at finite baryon densities. 
In this work, we estimate the energy-momentum tensor by including
constituent quarks. Technically it is done by counting
leading hadrons that have original constitute quarks within a formation time
multiplying the reduction factor of one-third (two-third) or one-half
depending on the number of constituent quarks inside baryons or mesons.
In the previous calculations in Ref.~\cite{Nara:2016phs},
we did not take into account the reduction factors
in leading hadrons which leads to larger energy densities.

\begin{figure}[t]
\includegraphics[width=8.0cm]{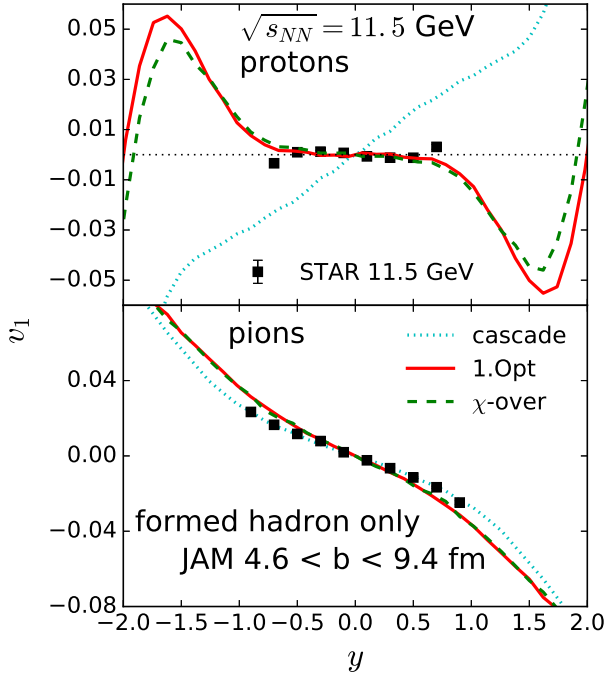}
\caption{Same as Fig.~\ref{fig:v1_11}, but only formed hadrons
are taken into account for the estimation of energy density.
}
\label{fig:v1_11had}
\end{figure}

To see the sensitivities of the different assumptions in defining
the energy density within our model, 
the proton directed flow from the calculation in which
only formed hadrons are taken into account for the estimation of
energy density~\cite{Sorge:1995pw}
 is presented in Fig.~\ref{fig:v1_11had}.
In this case, energy density corresponds to hadronic energy density, and
it is less than the default calculation,
in which the contributions from constituent quarks are included,
and as a result, we have more softening of EoS and the slope of proton directed
flow at $\srtNN=11.5$ GeV becomes negative.
However, we do not see any sensitivities
between the first order phase transition
and the crossover EoS at $\srtNN=11.5$ GeV.
We also note that pion directed flow from JAM deviates from the data 
when we use hadronic energy density.

\begin{figure}[tbh]
\includegraphics[width=8.5cm]{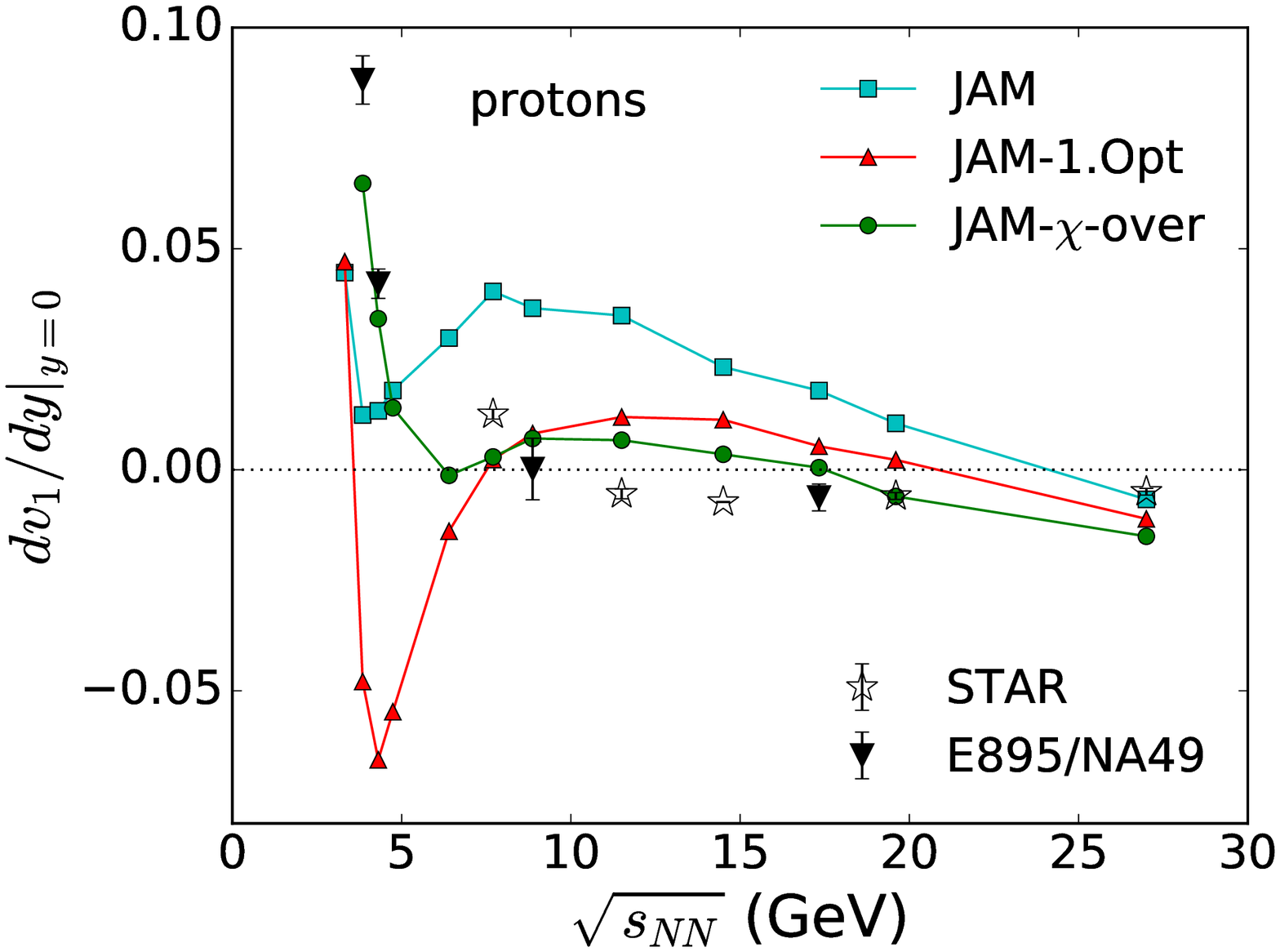}
\includegraphics[width=8.5cm]{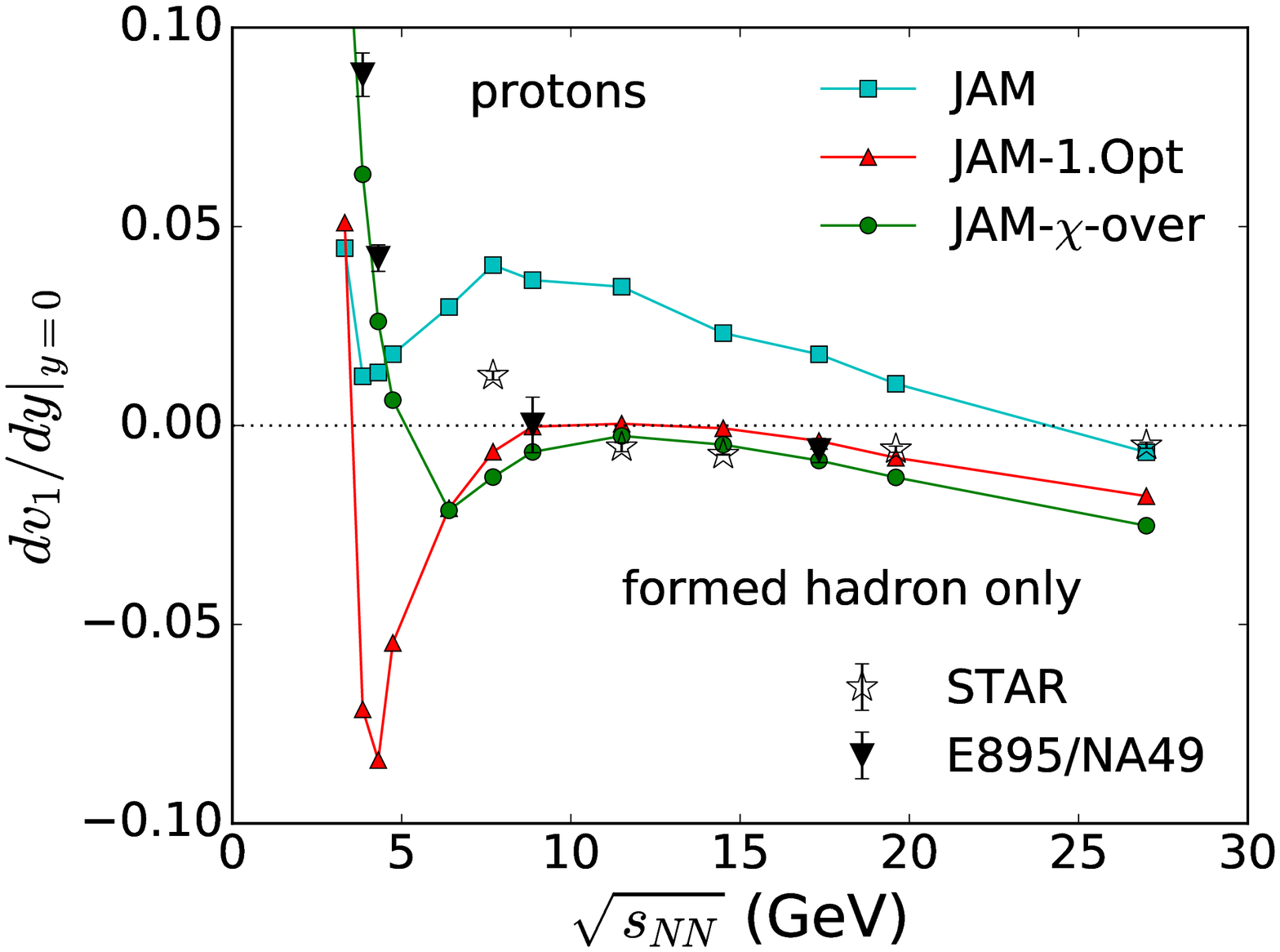}
\caption{Beam energy dependence of the slope of the directed flows of
protons in mid-central Au+Au collisions (10-40\%)
from JAM cascade mode (squares),
JAM with first-order EoS (triangles)
and with crossover EoS (circles)
in comparison with the STAR/NA49/E895
data~\cite{STARv1,Shanmuganathan:2015qxb,E895v1,NA49prl,NA49prc}.
Local energy densities are computed by taking into account
the contributions of constituent quarks in the upper panel,
while only formed hadrons are included in the lower panel.
}
\label{fig:v1slope}
\end{figure}
 
In the following we present 
the beam energy dependence of the slope of the directed flow.
In Fig.~\ref{fig:v1slope}
we show the slope of directed flow $dv_1/dy$
of protons in mid-central collisions from the standard JAM cascade and
JAM with a first-order phase transition and with a crossover EoS,
in Au+Au collisions
in comparison with the data from STAR, NA49 and E895 Collaborations
~\cite{STARv1,E895v1,NA49prl,NA49prc}.
The slope is obtained by fitting the rapidity dependence
of $v_1$ to a cubic equation $v_1(y)=Fy+Cy^3$
in the rapidity interval $-0.8<y<0.8$.
 
The standard JAM cascade calculation predicts a minimum at around
AGS energies which was first reported in Ref.~\cite{Bleicher:2002xx}
within the UrQMD approach. 
The decrease of the directed flow in the standard cascade
approach can be understood by the rapid change in degree of freedoms due to the excitation
of hadronic resonances up to 2 GeV.
Note that PHSD does not have such minimum
~\cite{Konchakovski:2014gda}, since there are only a few number of
hadronic resonances included in the PHSD model.
It is important to notice that the slope of the proton
directed flow obtained by the cascade model is
 still positive at the minimum point.
At higher energies up to $\srtNN\approx20$ GeV,
the JAM standard model overestimates the slope of
the proton directed flow as already reported in Ref.~\cite{Nara:2016phs}.

In the case of the EoS with the first order phase transition (JAM-1.Opt),
we also see the minimum in the excitation function of the proton directed
flow at almost the same beam energy as the cascade model.
In addition, the slope is now negative as predicted
by hydrodynamical approaches.
The beam energy dependence is very similar to the pure hydrodynamical simulation
except that the magnitude of the minimum in JAM-1.Opt is about a factor 5
smaller than the ideal hydrodynamical prediction~\cite{Steinheimer:2014pfa}, which may be related to
the finite viscosity in the transport approach.
The local minimum predicted by the JAM-1.Opt is located 
at a slightly higher beam energy than in the one-fluid model prediction ($\srtNN\approx4$ GeV)
in which the strong coupling of the fluids lead to an almost instantaneous full stopping and maximum energy deposition,
different than in the three-fluid model in Ref.~\cite{Brachmann:1999xt} in which
finite stopping power of nuclear matter is taken into account.
We note that the local minimum predicted by the three-fluid model
in Ref.~\cite{Ivanov:2014ioa}
shows $\srtNN=6.5$ GeV. Probably the location of minimum depends
both on the EoS and the degree of stopping and its modeling.
In Ref.~\cite{Ivanov:2014ioa}, the EoS in the QGP phase is modeled
by a quasi-particle approximation
 with mean field potential~\cite{Khvorostukin:2006aw}.
It is interesting to notice that the ART BUU approach with the first order
phase transition also exhibits minimum
with a negative slope~\cite{Li:1998ze} which supports that
our method effectively handles the effect of the EoS.
 
On the other hand,
in the case of a crossover EoS (JAM-$\chi$-over),
there is still a local minimum at $\srtNN\approx6$ GeV,
but it is not so pronounced.
Thus our approach supports the idea that
a large negative slope of the proton directed flow
would be a good observable to identify a strong softening in the early phase of the collision due to the first order phase transition.
However, the minimum observed from our approach is located at a beam energy similar to that predicted by
hydrodynamical simulations~\cite{Brachmann:1999xt} and much lower than the minimum measured by the STAR. 
If the EoS we employ is close to the true EoS in nature,
we do not see the connection between the softest point of 
the EoS and the STAR data.
In order to describe the minimum at a larger beam energy one would need a very soft EoS at very high baryon densities, much higher than
what is found in the currently used EoS, in order to shift the minimum to the higher beam energies.
Within our analysis, the reduction of the proton directed flow
at $\srtNN>10$ GeV
is essentially related to the early pressure in the pre-equilibrium 
stages of the reaction.
The reason why JAM-1.Opt yields stronger flow than
JAM-$\chi$-over at $\srtNN>10$ GeV
is simply because of the fact that JAM-1.Opt in the
current study assumes the massless ideal quark-gluon EoS at the QGP
phase, while crossover EoS is consistent with the lattice QCD data
at the vanishing baryon chemical potentials.
Even at the finite baryon chemical potentials, our crossover EoS
is softer than the massless ideal quark-gluon EoS at high energy densities.
Since a massless ideal QGP EoS is not supported
by the lattice QCD calculations, we need to test more realistic EoS 
in order to see the difference between a first order phase transition
and crossover transition in the future.

Also it has been pointed out that the usual description of baryon stopping
in string models in a transport model
assumes essentially an instant deceleration of the leading baryons
because strings are immediately decay into hadrons with a formation
time.
In \cite{Bialas:2016epd} it was argued that a constant deceleration would lead to a very different 
initial density profile in coordinate space,
i.e. a smaller density at mid-rapidity,
which may transform into a different time evolution of the particle flow. 

Besides an uncertainty of the EoS, there is a model uncertainty regarding
the handling of static equilibrium EoS $P(e)$ within a non-equilibrium
dynamical framework in which only effective local energy density
can be obtained.
To check how the directed flow is affected by
the different definition of the energy density,
we compute the slope of the proton directed flow by using hadronic
energy density by counting only formed hadrons in the estimation
of energy-momentum tensor which may be regarded as a possible
lower limit of the energy density in the model.
The results from JAM-1.Opt and JAM-$\chi$-over simulations
are shown in the lower panel of Fig.~\ref{fig:v1slope}.
Overall tendency does not change with this approach,
but the slopes of proton directed flow at higher beam energies
becomes negative compatible with the data up to $\srtNN=19.6$ GeV
except the point at $\srtNN=7.7$ GeV.
We demonstrate that the directed flow of proton is highly sensitive
to the EoS at the beam energy range of $4\leq\srtNN\leq10$ GeV,
where no experimental data exists between 4.7 and 7.7 GeV.
It is utmost important to measure the directed flow in this range 
in order to reveal the phase structure of QCD at the highest baryon densities.

In summary,
we have investigated the EoS dependence of the excitation function
of the directed flow of protons within a microscopic
transport approach by employing an energy density dependent 
EoS modified collision term.
We showed that our approach provides an efficient method to control the
EoS in a microscopic transport model,
and our result yields qualitatively similar result as the pure hydrodynamical predictions.
As predicted by hydrodynamical approach,
the minimum of the directed flow, from a strong first order phase transition, is located at much lower
beam energy than the STAR data.
In order to distinguish between a first order phase transition
and crossover transition in the directed flow data, we need to examine 
different EoS from models consistent with the lattice QCD up to
finite baryon chemical potentials accessible in the current lattice QCD
calculations.
A systematic study including strange hadrons
should be addressed since strange hadrons follow a more complex
pattern due to the different cross sections~\cite{Nara2017}.
In the future, it is interesting to analysis STAR data
by explicitly implementing the mean field with phase transition
into a transport model.
It is also interesting to investigate the effect of dropping of hadron
mass due to CSR on the directed flow.

We suggest to study the directed flow in the beam energy
region of $4.7<\srtNN<11.5$ GeV which may give suitable signatures to
study the properties of EoS at highest baryon density, where
the softening effect could be best manifested.
Future experiments such as the BES II at RHIC~\cite{BESII},
FAIR~\cite{FAIR},
NICA~\cite{NICA},
and J-PARC~\cite{Sako:2014fha}
should provide important information on the properties of high
dense matter created in the heavy ion collisions.

\section*{Acknowledgement}
Y. N. thanks the Frankfurt Institute of Advanced Studies where part of this
work was done.
This work was supported in part by 
the Grants-in-Aid for Scientific Research from JSPS
(Nos.
 15K05079
and
 15K05098 
),
H.~N.~ has received funding from the European Union's Horizon 2020 research and
innovation programme under the Marie Sklodowska-Curie grant agreement
no. 655285 and from the Helmholtz International Center for FAIR
within the framework of the LOEWE program launched by the State of Hesse.
Computational resources have been provided by
the Center for Scientific Computing (CSC) at the Goethe-University of Frankfurt
and GSI (Darmstadt).

\end{document}